\documentstyle[twoside,fleqn,espcrc2]{article}

\newcommand{\AmS}{{\protect\the\textfont2
  A\kern-.1667em\lower.5ex\hbox{M}\kern-.125emS}}

\newcommand{\ssi}{\mbox{$\langle\overline{\psi}\psi\rangle$}}

\title{Topology in QCD}

\author{M. Teper \\
\vspace{0.1in}
        Theoretical Physics, University of Oxford, \\
        1 Keble Road, Oxford OX1 3NP,  United Kingdom}
       
\begin{document}


\maketitle

\section{Introduction}

Topology is important in QCD largely due to
the way it influences the propagation of quarks.
Here it is the zero-modes of {\mbox{$i\!\!\not\!\!D$}}
that are special; not the exact zero modes
whose number per unit volume 
\begin{equation}
{{\overline{|Q|}}\over{V}} \propto {1\over{\sqrt{V}}}
\stackrel{V\to\infty} \longrightarrow 0,
\end{equation}
but rather the (mixed) would-be zero modes that 
would have been exact zero-modes if 
their parent topological charges had been isolated.
The total number of these is certainly $\propto V$,
if only from the small instantons whose density is 
calculable analytically. Lattice calculations find,
in both SU(2) and SU(3), a substantial value for
\begin{equation}
{{\langle Q^2 \rangle}\over{V}} 
\sim (200 \ MeV)^4 
\simeq {1\over{fm^4}} 
\end{equation}
and this suggests that alot of the topological 
charge density tends to cluster in lumps that are
uncorrelated at larger distances -- just like a
`gas' of instantons. So, for convenience, this
is the language I shall use in this talk.

I will start with the aspect that is no doubt on the firmest
theoretical ground: the large mass of the $\eta^{\prime}$
\cite{thooft} and the value \cite{wit-ven}
of the topological susceptibility
$\chi_t \equiv \langle Q^2 \rangle/V$. I dwell on the
quenched case because it provides one area in which the 
lattice calculations are in very good shape, even if 
there is not much new this year (probably because
they are in such good shape ...). I will then move
onto the possible role of instantons in driving
the spontaneous breaking of chiral symmetry
\cite{oldssb,diakonov,shuryak}.
Here the exciting news is that we have finally got 
lattice fermions that are good enough to address
this question realistically. More speculative is
the influence of topology on hadrons
\cite{shuryak}. 
This provides a motivation for trying to extract the 
properties of the `gas' of instantons in the vacuum. 
There has been some interesting  new work on the
latter during the past year. Finally I
turn to the role of instantons in confinement.
The usual view is that there is no such role.
I will discuss a recent calculation that claims
the opposite. This will be the one area where
fermions don't appear at all.

\section{$Q$, $\chi_t$ and $m_{\eta^{\prime}}$}

Instantons provide a resolution of the 
$\eta^{\prime}$ puzzle \cite{thooft}. The mass of
the $\eta^{\prime}$ can be related to the strength
of topological fluctuations \cite{wit-ven}:
\begin{equation}
\chi_t \equiv 
{{\langle Q^2 \rangle}\over{V}} 
\simeq {{m^2_{\eta^{\prime}}f^2_{\pi}}\over{2N_f}}
\sim (180 \ MeV)^4. 
\label{eqn-wv}
\end{equation}
This is to leading order in $N_c$ and so the value
of $\chi_t$ is the quenched value. The practical
application (i.e. $MeV$ units) assumes that $N_c=3$
is close to $N_c =\infty$ and this is also
something that one needs to confirm.

Let me start with some reassuring comments about calculating
the topological charge $Q$ of lattice gauge fields.
We suppose the lattice spacing $a$ is very small. Consider 
topological charges which are localised within a core of
radius $\rho$. For $a \ll \rho \ll 0.5fm$ these charges
have an analytically calculable density, which turns out to 
be $\propto \rho^6$ for SU(3). Large scales, say 
$\rho \geq 0.5fm$, are a non-perturbative problem. 
At the ultraviolet scale, $\rho \sim a$, lattice artifacts
may dominate. Now, the point to note is that during the Monte 
Carlo each step changes only one link matrix i.e. the fields 
within a volume $\delta v \sim a^4$. So the only way we can 
change the value of $Q$ is for the core of the 
topological charge to shrink over many Monte Carlo 
steps from say $\rho \sim 0.2 fm$ down to $\rho \sim a$ 
and then within a hypercube and so out of the lattice.
Now because the density $\propto \rho^6$, the region
$a \ll \rho \ll 0.1fm$ (say) will normally (in our
finite lattice volume of a couple of fermi) have no
instantons at all. So as an instanton traverses this
topological desert it will become very visible --
particularly once $\rho \sim few\times a$. At this point
the core will stick out above the $O(1/\beta)$ UV fluctuations
in both the action and the topological charge densities. 
Now, if we know when $Q$ changes then that is all we need 
to calculate ${\langle Q^2 \rangle}$ (in a long run); and
we can calculate $Q$ e.g. by averaging over long sequences 
of configurations. Clearly the same argument applies to
cooling, and because cooling rapidly dampens the UV fluctuations,
the above scenario can work for modest $a$ with only a couple of
cooling sweeps being performed on each Monte Carlo field in
the sequence.

So as $a\to 0$ we can certainly calculate $Q$ and 
 ${\langle Q^2 \rangle}$ for lattice gauge fields. 
At finite-$a$ there will typically be $O(a^2)$ corrections
whose size will depend on the method used. Note that things 
will be worse in SU(2) than in SU(3). In SU(2) the  
density of small instantons
decreases only as $\propto \rho^{7/3}$: that is to say, the 
decoupling of the UV scale occurs much more slowly. Although
one often starts with SU(2) calculations because they are much 
faster, I think that, for the above reason, this is a false economy.

I will now summarise a number of different calculations of 
$\chi^{1/4}_t/\surd\sigma$. If we use the usual form of the 
lattice topological charge \cite{oldffd} 
\begin{equation}
Q_L = {1\over{32\pi^2}} \sum_x \varepsilon_{\mu\nu\rho\sigma}
Tr\{U_{\mu\nu}(x)U_{\rho\sigma}(x)\}
\end{equation}
(then $\pm\mu$ etc. antisymmetrised)
and if we average $Q^2_L$ over fields containing a topological 
charge $Q$ we obtain
\begin{equation}
{\overline{Q^2_L}} = Z^2(\beta)Q^2 + \zeta(\beta).
\label{hotQ}
\end{equation}
These additive \cite{oldffd} and multiplicative \cite{pisaZ} 
renormalisations can be either calculated analytically
in powers of $1/\beta$, or numerically\cite{oldffd,pisaZ,myZ}. 
(Note that although
$\zeta(\beta)$ diverges as $1/a^4$, in practical calculations
it is not much larger than $Q^2$ and so the `subtraction'
is under much better control than in the
corresponding gluon condensate calculations.) Alternatively 
one can cool the fields, and then $Z \to 1$, up to $O(a^2/\rho^2)$ 
lattice corrections, and $\zeta \to 0$, after a couple of cools.

I now take a number of quite different lattice calculations
(using only those with at least 3 $\beta$ values)
and compare their continuum extrapolations. I extrapolate
the dimensionless ratio $\chi^{1/4}_t/\surd\sigma$
($\sigma$ = string tension) 
\begin{equation}
{{\chi^{1/4}_t(a)}\over{\sqrt\sigma(a)}} =
{{\chi^{1/4}_t(0)}\over{\sqrt\sigma(0)}} + c a^2\sigma
\end{equation}
using a common set of values for $\sigma$ 
(as listed in \cite{myK}) so that any differences are 
differences in the calculation of topology, not of the scale.
I now list the SU(3) calculations I use, and the results 
of the corresponding continuum extrapolations.

{\noindent}$\bullet$ {\bf SU(3)}

{\noindent}{\it Pisa}\cite{pisasu3} Smeared version of
$Q_L(x)$, calculated directly from Monte Carlo field
average using eqn~\ref{hotQ} 
($5.9 \leq \beta \leq 6.1$): 
$\chi^{1/4}_t/\surd\sigma = 0.464(23)$

{\noindent}{\it Boulder}\cite{bouldersu3} Algebraic
lattice $Q$ on RG smoothed fields
($5.85 \leq \beta \leq 6.1$): 
$\chi^{1/4}_t/\surd\sigma = 0.456(23)$

{\noindent}{\it Oxford}\cite{oxfordsu3} $Q_L$ 
on cooled fields
($5.7 \leq \beta \leq 6.2$): 
$\chi^{1/4}_t/\surd\sigma = 0.449(17)$

{\noindent}{\it UKQCD}\cite{ukqcdsu3} $Q_L$ 
on cooled fields
($6.0 \leq \beta \leq 6.4$): 
$\chi^{1/4}_t/\surd\sigma = 0.448(50)$

{\noindent}$\bullet$ {\bf SU(2)}

{\noindent}{\it Pisa}\cite{pisasu2} Smeared version of
$Q_L(x)$, calculated directly from Monte Carlo field
averages using eqn~\ref{hotQ} 
($2.44 \leq \beta \leq 2.57$): 
$\chi^{1/4}_t/\surd\sigma = 0.480(23)$

{\noindent}{\it Boulder}\cite{bouldersu2} 
lattice $Q$ on RG mapped fields
( $2.4 \leq \beta \leq 2.6$): 
$\chi^{1/4}_t/\surd\sigma = 0.528(21)$

{\noindent}{\it Oxford-Liverpool}\cite{ox-livsu2} 
 $Q_L$ on cooled fields
( $2.2 \leq \beta \leq 2.6$): 
$\chi^{1/4}_t/\surd\sigma = 0.480(12)$

{\noindent}{\it Oxford}\cite{oxfordsu2} 
blocked geometric $Q$ directly on Monte Carlo fields
($2.3 \leq \beta \leq 2.6$): 
$\chi^{1/4}_t/\surd\sigma = 0.480(18)$

{\noindent}{\it Zurich}\cite{zurichsu2} 
version of $Q_L$ on improved-cooled fields
($2.4 \leq \beta \leq 2.6$): 
$\chi^{1/4}_t/\surd\sigma = 0.501(45)$

\vskip 0.1in

All these continuum limits are consistent
with each other despite the wide variety of
methods being used. They provide the following 
conservative estimates:
\begin{equation}
{{\chi^{1/4}_t}\over{\surd\sigma}} 
 = \cases {
0.455 \pm 0.015 &$ \ \ \ \ SU(3)$ \cr
0.487 \pm 0.012 &$ \ \ \ \ SU(2)$. \cr} 
\end{equation}
We observe that the two values are close to
each other suggesting that  (for the
pure gauge theory) SU(3) is indeed close
to SU($\infty$). If we plug in the value
${\surd\sigma} \simeq 440 \pm 38 MeV$ \cite{myK},
then we find 
$\chi^{1/4}_t = 200 \pm 18 MeV$ for SU(3).
As we have seen this is roughly the density of 
topological fluctuations that is needed \cite{wit-ven}
to provide the $\eta^\prime$ with its observed mass.

One can of course also calculate $\chi_t$ for `full QCD'.
We expect that $\langle Q^2 \rangle \to 0$ as $m_q \to 0$,
since the zero modes ensure that 
$\det${\mbox{$i\!\!\not\!\!D$}}=0 for $Q\not= 0$.
In fact we know more: the anomalous Ward identities tell us that
\begin{equation}
\chi_t = {{m^2_\pi f^2_\pi}\over{n^2_f}} + O(m^2_q) \propto m_q
\label{chi-dyn}
\end{equation}
if we are in the phase in which chiral symmetry is
spontaneously broken. By contrast, in a chirally symmetric
phase we expect $\chi_t \propto m^{n_f}_q$, and, of course, in
the quenched case  $\chi_t \propto m^0_q$.  So we can test the 
lattice calculations against this relation. One has to be 
careful because the lattice spacing $a$ is both a function
of $\beta$ and $m_q$. (For a large quark mass the running
of the coupling will include the quark only for scales 
below $O(1/m_q)$.) So in testing eqn~\ref{chi-dyn} we should 
express all the quantities in terms of some physical
quantity that is not expected to vary strongly with $m_q$, 
e.g. $r_0$ or $\surd\sigma$. The ($n_f=2$) UKQCD calculations
\cite{ukqcd-dyn} reported at this meeting go one better
by tuning $\beta$ with $m_q$ so that $r_0/a$ is independent
of $m_q$. Such a calculation separates the $m_q$ dependence
from any $a$ dependence. If one plots $r^4_0\chi_t$ versus
$r^2_0 m^2_\pi \propto m_q$ one finds \cite{ukqcd-dyn}
that the (three)
points are consistent with eqn~\ref{chi-dyn} but only
if one decreases $f_\pi$ by about 20\% from its
physical value. That, I think, is pretty good. Less
pretty are the large statistical errors. Despite the latter
it is clear that the value of $r^4_0\chi_t$ has decreased
from its quenched value and that a $\propto m^{n_f=2}_q$
behaviour is excluded. This contrasts with the $n_f=2$ 
CP-PACS calculation \cite{cp-pacs} of $\chi_t/\sigma^2$
which strangely finds no $m_q$ dependence at all. On the
other hand the $n_f=2$ calculation of the Pisa group
\cite{pisa-dyn} does show some sign of a decrease as
$m_q \to 0$.

The meson that (perhaps) most directly reflects
topological fluctuations is the $\eta^\prime$.
CP-PACS has produced \cite{cp-pacs} a very nice 
calculation of  $m_{\eta^\prime}$. This is a 
tough calculation because such flavour-singlet
hadrons  simultaneously need the statistics of glueball 
calculations and
expensive quark propagators. CP-PACS do a direct
calculation that finds $m_{\eta^\prime}\not= 0$ 
as $m_q \to 0$, in contrast to the $\propto m_q$ 
Goldstone behaviour one would expect if topological 
fluctuations were negligible. The value they
obtain in the continuum chiral limit is 
$m_{\eta^\prime}=863(86)MeV$.
This is for $n_f=2$ and $n_c=3$, so it is amusing to note 
that if we plug into eqn~\ref{eqn-wv} the values
$n_f=2$, $f_\pi=93MeV$ (since this should be
insensitive to the strange quark) and $\chi_t=(200MeV)^4$ 
(the lattice value) then we obtain 
$m_{\eta^\prime} \simeq 860MeV$ as our expectation.
This is promising and I hope CP-PACS will pursue
this calculation; perhaps in the direction 
\cite{ven-kilcup} of explicitly showing that
it is dominated by the lowest modes of 
{\mbox{$i\!\!\not\!\!D$}}. A related calculation
has been performed by UKQCD \cite{ukqcd-eta};
and the Pisa group \cite{pisa-priv} has tried 
calculating the mass using correlators of
the ($\vec{p}=0$) topological charge.

\section{Chiral symmetry breaking}

Let $\rho(\lambda)$ be the normalised spectral density
of {\mbox{$i\!\!\not\!\!D[A]$}} averaged over gauge
fields. Then we can express the chiral condensate as
\begin{eqnarray}
\ssi 
& = & 
\lim_{m \to 0} \lim_{V \to \infty} \ssi_{m,V} \nonumber\\
& = &  \lim_{m \to 0} \lim_{V \to \infty}
\int_{0}^{\infty}\frac{2m\rho(\lambda,m)}{\lambda^{2} + m^{2}}  
d\lambda \nonumber\\
& = & 
\pi\rho(0)
\end{eqnarray}
So chiral symmetry breaking requires a non-zero density of
modes at $\lambda=0$. Now if we remove all interactions, then
each (anti)instanton has a zero-mode and so 
$\rho(\lambda) \propto \delta(\lambda)$. Interactions will spread 
these modes away from zero: however instantons are clearly a good 
first guess if what you want is $\rho(0)\not= 0$. Contrast it
with the non-interacting limit of the perturbative vacuum
-- the free theory -- where  $\rho(\lambda) \propto \lambda^3$.

This idea, that instantons might drive chiral symmetry breaking, is 
an old one \cite{oldssb}. There have been lattice calculations to test
this idea, for example \cite{shmt}. The calculations in \cite{shmt},
with staggered fermions in the pure gauge SU(2) vacuum, found
that the chiral symmetry breaking disappeared if one removed
the topological eigenmodes of {\mbox{$i\!\!\not\!\!D[A]$}}.
While conclusive about the lattice theory, there was a question
mark over the continuum theory: despite the lattice
spacing being `small' (e.g. $\beta=2.6$) the $|Q|$ zero-modes 
were no closer to zero than the other small modes of
{\mbox{$i\!\!\not\!\!D[A]$}}. (See also
\cite{9909017}.) This raises doubts about
how continuum-like is the spectrum of the $O(V)$ mixed
would-be zero modes; one really needs the lattice shift in the
exact zero-modes to leave them small compared to the other small 
modes in a reasonably sized box. (The exact $|Q| \sim \sqrt{V}$ 
zero modes become irrelevant in the thermodynamic limit.)
This raises doubts about any lattice calculations
of the influence of instantons on quarks; e.g hadron masses
as well as chiral symmetry breaking.

So the exciting news here is, of course, `Ginsparg-Wilson'.
In particular the Columbia group has pursued the domain
wall variant and have produced some very pretty
calculations \cite{columbia} showing that even with a modest 
5'th dimension one obtains `exact' lattice zero-modes
that are very much smaller than the other small modes
of the Dirac spectrum in a reasonably sized box. 
(In practice they observe the $\propto 1/m$ behaviour
in $\ssi_{m,V}$ that one obtains from such modes.)
This provides an explicit demonstration that controlled
calculations of the influence of topology on continuum
hadronic physics can now be done.

There has also been related work with overlap fermions
\cite{overlap} and there has been a great deal of comparison 
with Random Matrix Theory, as well as other model calculations, 
which I hope will be reviewed elsewhere.

\section{The instanton content of the vacuum}

There has been a new calculation of the topological structure 
of the SU(2) vacuum \cite{perez}. The main novelty here is
a modified cooling algorithm that is designed to cool out
to a specified cooling radius $r_c$. One finds that while
some features, such as the average instanton size, vary weakly
with  $r_c$, other quantities, such as the instanton density
and average nearest instanton anti-instanton distance,
vary rapidly with  $r_c$. (Not unlike the conclusions of 
\cite{dsmt} using ordinary cooling.) 
This is largely bad news if what 
you want is to obtain the detailed properties of the
instanton `gas' in the original vacuum, so that you can
provide an input into phenomenological instanton 
calculations \cite{shuryak}. 

However one thing that is not understood is how much of this 
apparent variation with cooling is real and how much of it
is actually the fault of the `pattern recognition'
algorithms that turn the topological charge density
into an instanton ensemble. Here a gleam of hope
comes from a reanalysis in \cite{arfs} of \cite{dsmt}.
Clearly in $n_c$ usual cooling sweeps one expects roughly 
$\overline{r_c} \propto a \sqrt{n_c}$. Plotting the
data of \cite{dsmt} for all $\beta$ and $n_c$ against
the corresponding scaling variable, \cite{arfs} finds
a very nice scaling. More importantly, they find a
range of $r_c$ where the instanton properties become
independent of $r_c$. This range is narrow, so it needs
more work. But it provides some hope ...

The instanton gases found on the lattice are usually
denser, and with larger average instanton sizes, than
the instanton liquid models \cite{shuryak} would like.
So it is interesting that a recent calculation \cite{usmt} 
of the quark physics (in a `toy model') from the lattice
instanton ensembles of \cite{dsmt} finds that the 
important part of the low-$\lambda$ spectrum looks like
that of a dilutish gas of narrower instantons. There
is a simple reason for this; a large instanton 
has a large zero-mode that has a correspondingly
small density. It will therefore have a small
value when integrated over the small volume where
the zero-mode of the small instanton resides.
This leades to a small mixing between large and
small instantons: they approximately decouple.
This mechanism provides a possibility for reconciling the
lattice and the instanton liquid.

Note also related work in \cite{kovacs}
and \cite{annah}.

\section{Instantons and confinement}

An isolated instanton affects a large Wilson loop weakly;
it merely renormalises the Coulomb potential \cite{oldssb}.
So one expects that an instanton `gas' (no long range order) 
will not disorder a Wilson loop strongly enough to produce
an area decay. The claim in \cite{fst} that random ensembles
of instantons do produce linear confinement is therefore
surprising. However we note that our analytic intuition 
really holds for Wilson loops that avoid instanton cores.
In  \cite{fst} the density distributions used are
$D(\rho) \propto 1/\rho^5$ or $1/\rho^3$. This corresponds
to the `packing fraction' $\propto \int d\rho D(\rho) \rho^4$
diverging! So these are very dense gases, and the Wilson loop 
will, throughout its length, pass in the middle of densely
overlapping instanton cores. It may be that this will
disorder the Wilson loop sufficiently to confine. This is
relevant since lattice calculations suggest that instantons
in the real vacuum are dense and highly overlapping. It
is of course important to check that the field configurations
being used, generated by the approximation of adding the  
individual $A_I(x)$ (in singular gauge), do indeed accurately 
denote a `gas' of topological charges, and that the
confining disorder is not being produced by the breakdown of 
this standard linear addition ansatz. These and related
\cite{fst,conf} directions will be interesting to pursue further.

\end{document}